\documentclass[12 pt]{article}
\usepackage{graphicx}

\begin{document}

\centerline{\bf \large
Properties of the energetic particle distributions}
\centerline{\bf \large during the October 28, 2003 
solar flare}
\centerline{\bf \large from INTEGRAL/SPI observations}
\vspace{1 cm}

\normalsize
\centerline{J. Kiener$^1$, M. Gros$^2$, V. Tatischeff$^1$, and
G. Weidenspointner$^3$}  
\scriptsize
\centerline{(1) C.S.N.S.M., IN2P3-CNRS et Universit\'e Paris-Sud, 91405 Campus
Orsay, France} 

\centerline{(2) DSM/DAPNIA/Service d'Astrophysique, CEA Saclay, 91191 Gif-sur-Yvette,
France} 

\centerline{(3) Centre d'Etude Spatiale des Rayonnements, 9, avenue du Colonel Roche, 31028
Toulouse, France}  

\abstract{ 

Analysis of spectra obtained with the gamma-ray spectrometer SPI onboard
INTEGRAL of the GOES X17-class flare on October 28, 2003 is presented. In the
energy range 600 keV - 8 MeV three prominent narrow lines at 2.223, 4.4 and 
6.1 MeV, resulting from nuclear interactions of accelerated ions within the
solar atmosphere could be observed.  Time profiles of the three lines and the
underlying continuum indicate distinct  phases with several emission peaks and
varying continuum-to-line ratio for several minutes before a smoother decay phase
sets in. Due to the high-resolution Ge detectors of SPI and the exceptional
intensity of the flare, detailed studies of the 4.4 and 6.1 MeV  line shapes
was possible for the first time.  Comparison with calculated line shapes using
a thick target interaction model and several energetic particle angular
distributions indicates that the nuclear interactions were induced by 
downward-directed particle beams with alpha-to-proton
ratios of the order of 0.1. There are also indications that the 4.4 MeV to 6.1
MeV line fluence ratio changed between the beginning and the decay phase of the
flare, possibly due to a temporal  evolution of the energetic particle
alpha-to-proton ratio. 
 
{Sun: flares; Sun: gamma rays; gamma rays: 
observations; gamma rays: line profiles } }

\section{Introduction}

Gamma-ray lines from nuclear interactions in solar flares contain important 
information on the properties of the accelerated particle distributions and the
solar atmosphere. The relative line intensities are sensitive to the
composition and energy spectrum of  the energetic particles and the composition
of the  interaction region. Position and width of the lines put further
constraints on  the particle-momentum distribution, and may reveal in
particular the angular  distribution of the energetic particles. Line shape
calculations for solar  flares were pioneered by Ramaty et al. \cite{ramaty}
and subsequently developed  (Murphy et al. \cite{mur88}, Werntz et al.
\cite{werntz}, Kiener et al. \cite{kiener}) and applied to several flares with
strong  nuclear lines (Murphy et al. \cite{mur90},  Share \& Murphy
\cite{share97}, Share et al. \cite{share02}).  However, most of the solar flare
gamma-ray data available then were obtained by  satellites equipped with
scintillation detectors, which provided only moderate  energy resolution and
therefore could not exploit the  full potential of line shape  analysis.

Since the launch of RHESSI and INTEGRAL in 2002, both equipped with 
high-resolution gamma-ray cameras, spectra from several solar flares on 23
July  2002 (Smith et al. \cite{smith}, Share et al. \cite{share03}) and in
particular from the period  of extreme solar activity in late October/early
November 2003 (Gros et al. \cite{gros},  Share et al. \cite{share04}) have been
obtained which offer the opportunity to apply detailed line  shape analysis
techniques. We concentrate here on the spectra of the GOES class X17.2 flare of
October 28, 2003 obtained with the spectrometer SPI on board INTEGRAL. These
spectra cover the totality of the flare which lasted roughly 15 min in the
gamma-ray  band. The flare exhibited three different phases, first a continuum
emission without  notable lines during the first minute before the onset of
strong nuclear line  emission in a second phase lasting about three minutes and
third, a slow decline of line  and continuum component. 

Despite the unusual incidence of photons in SPI from a direction almost
perpendicular to the normal viewing direction, the  lines at 4.4 MeV from
$^{12}$C deexcitation and at 6.1 MeV from $^{16}$O deexcitation  could be
obtained with good statistics for the second and third phase of the  flare.
Given the exceptional intensity of  the flare in gamma rays and the fact that
RHESSI missed the major  part of the gamma-ray emission of this event, the two
strongly red-shifted and  broadened high-energy lines observed by SPI are
certainly the best case for  detailed line shape studies yet available.

Several properties of the accelerated particle distributions were
investigated,  using relative line intensities and the line profiles.  
Comparison of the observed line profiles at 4.4 and 6.1 MeV with calculated 
ones employing a thick-target interaction model and recent nuclear physics data
and calculations for  the production of both lines by the most important
nuclear interactions provided  interesting constraints on the alpha-to-proton
ratio ($\alpha$/p) and the geometry of the  energetic particle angular
distributions in the solar atmosphere.

\section{Observation}

\begin{figure} 
\resizebox{\hsize}{!}{\includegraphics{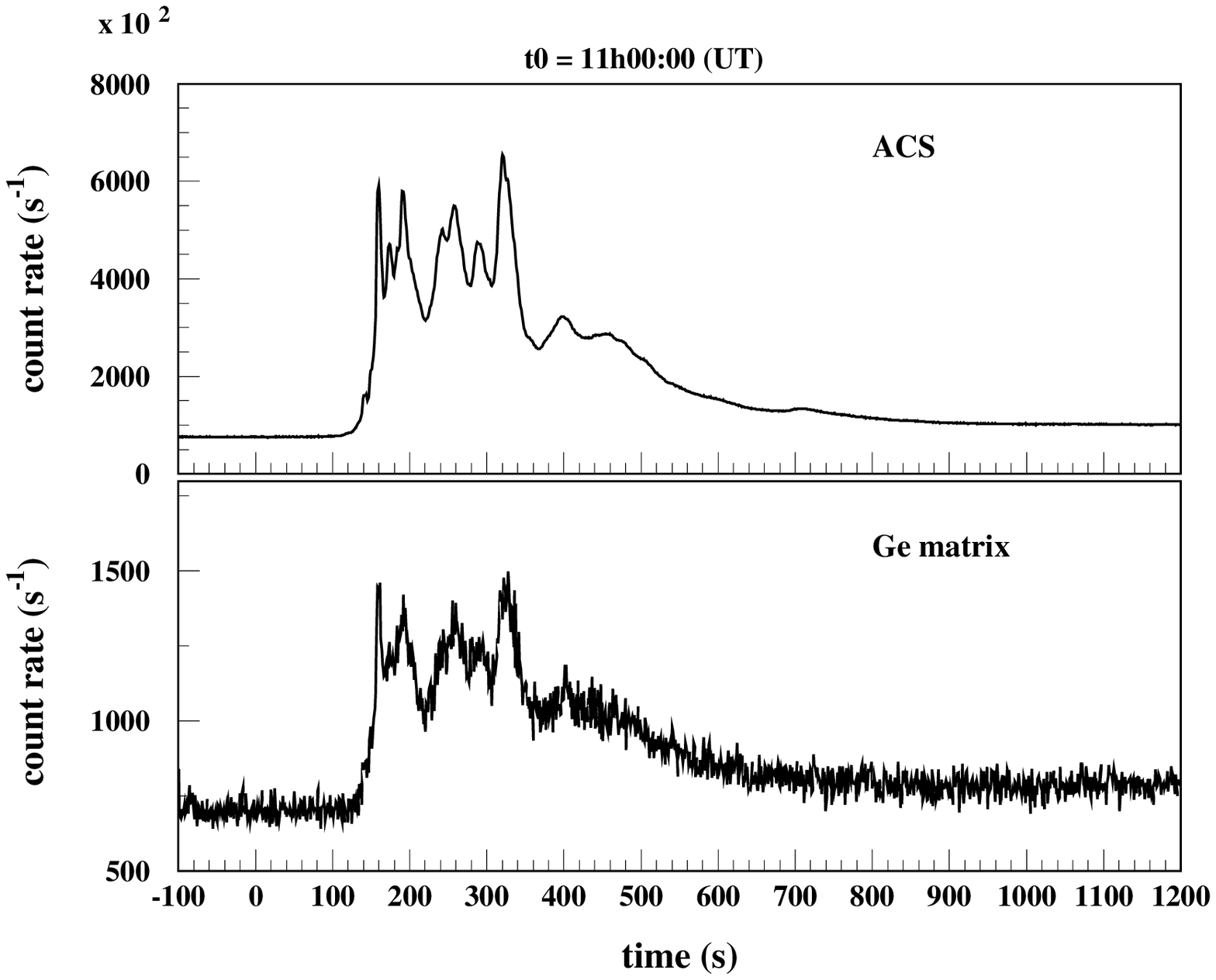}} 
\caption{
Total count rates of the BGO anti-compton shield (ACS) of SPI and of the SPI
Ge-detector array (Ge matrix) in 1 s resolution. Both count rates are corrected
for dead time, which attains 33\% as a maximum for ACS and 35\% for the Ge
matrix.  In the particular observation conditions for this flare, ACS and 
the Ge  matrix are sensible to photons above $\approx$ 150 keV and above 
$\approx$ 500 keV, respectively.}
\label{fig1}
\end{figure}

As observed by the GOES satellites, the X-ray flux of the 28 October 2003 solar
flare began at 9:41  UT at solar coordinates 18E20S, had its maximum at  11:10
and ended around 11:24 UT.  At that time INTEGRAL was pointed to the supernova
remnant IC443, with  the { Sun} at 122$^{\circ}$ from the instrument axis. The
flare was detected by the main  instruments of INTEGRAL, IBIS and SPI and their
BGO anti-coincidence shieldings. The most sensitive instruments in this
particular configuration were the shieldings because of their large area and
the relatively small absorption of the  photons in structural material before
reaching the scintillation crystals. In particular the BGO anti-coincidence
shield of SPI showed a strong increase of the counting  rate, starting slightly
after 11h02 with several narrow peaks during the most intense phase before a
somewhat smoother decay phase set in (fig.~\ref{fig1}). The behaviour is
similar  for the total count rate in the Ge-detector matrix but with a smaller
amplitude due to the strong additional absorption occuring mainly in the 
$\approx$5 cm of BGO crystals, which the solar gamma rays had to pass before
hitting the Ge matrix. The radiation background from high-energy particles
during the  gamma-ray flare was on a very low and constant  level as observed
by the GOES satellites and the particle monitor IREM onboard INTEGRAL,
providing good observing conditions in particular for the nuclear deexcitation
lines in the Ge detectors. Details of the observation conditions are given in
Gros et al. \cite{gros}. 																																																																																																																																			    

\begin{figure} 
\resizebox{\hsize}{!}{\includegraphics{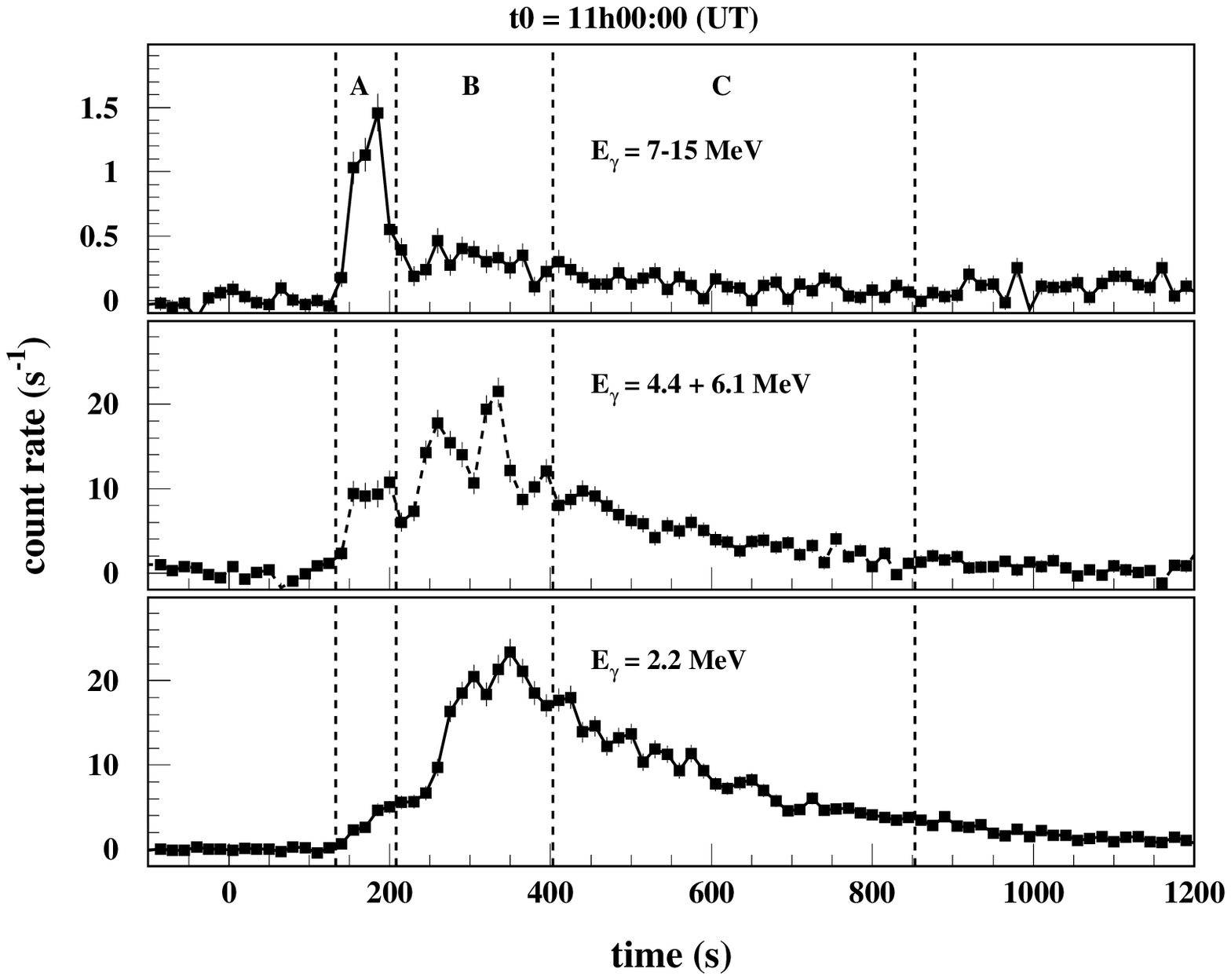}} 
\caption{
Dead-time corrected 15 s time profiles in different energy bands. 
Count levels
during the quiet period before the flare have been subtracted. The profile for
the continuum band 7.5 - 15 MeV has been obtained by simple integration while
the profiles representing the sum of the  strongest two 
nuclear deexcitation lines of 
$^{12}$C at 4.4 MeV and $^{16}$O at 6.1 MeV and the neutron-capture line at 2.223
MeV have been obtained by subtracting a continuum component estimated from
count rates below and above the respective lines. Phases marked A, B and C
define the time intervals 133-208, 208-403 and 403-853 seconds after 11h00 UT,
respectively.} 
\label{fig2}
\end{figure}

Due to the strong absorption in this configuration, the effective detection
thresholds of the BGO shield and the Ge matrix are respectively $\approx$150
keV and $\approx$500 keV. Several narrow lines at 2.223 MeV from neutron
capture on hydrogen, at 4.4 MeV from $^{12}$C,  6.1, 6.9 and 7.1  MeV from
$^{16}$O, 1.37, 1.63 and 1.78 MeV from $^{24}$Mg, $^{20}$Ne and $^{28}$Si
deexcitation and a continuum component extending up to at least 8 MeV could  be
observed in the Ge detectors. Time profiles of the neutron capture line, the
strongest two deexcitation lines and the continuum component are shown in
fig.~\ref{fig2}. They were obtained by subtracting the average count rate in
the respective energy range from the period just before the onset of the flare
and by subtracting additionally a  continuum component from the lines,
estimated from count rates in the region  above and below the line. Three
different phases can be distinguished in these  curves. In the first phase
lasting slightly more than a minute  (named phase A), a nearly line-free
bremsstrahlung continuum is observed, followed by strong nuclear line emission 
during several minutes (phase B) before a slow decline of line and continuum 
emission sets in (phase C). 

\begin{figure} 
\resizebox{\hsize}{!}{\includegraphics{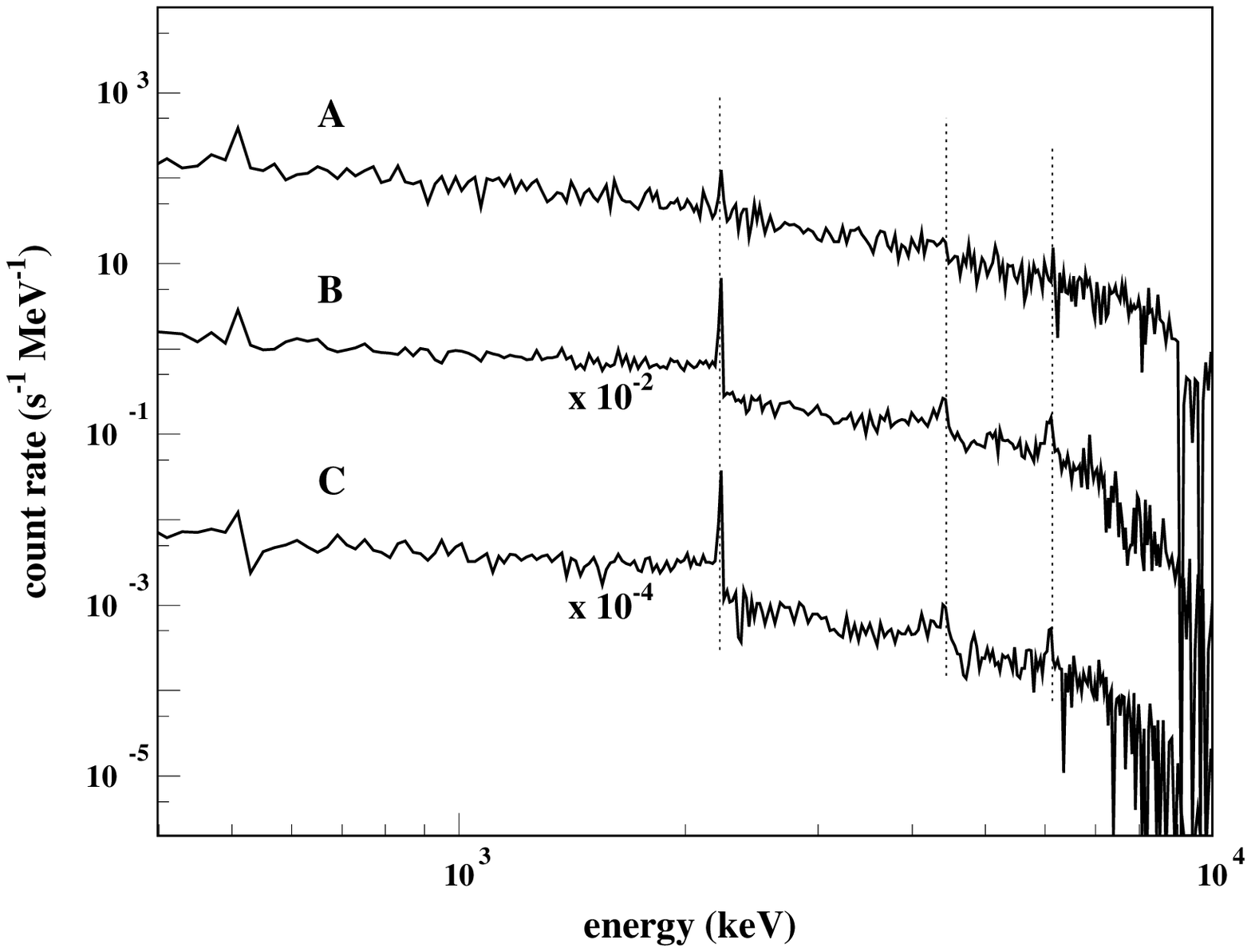}}
\caption{
Background subtracted spectra of time intervals A, B and C. The dotted lines
mark the nominal energies of the 2.223, 4.4 and 6.1 MeV lines.  }
\label{fig3}
\end{figure}

Count rate spectra for the different phases were obtained by subtraction of a
background spectrum which was obtained from a time interval of 15 minutes  just
before the onset of the gamma-ray flare. The normalization factor for the
subtraction was obtained from the  relatively
intense background lines at 139.7 and 198.4 keV from isomeric decays of
$^{75m}$Ge and $^{71m}$Ge, respectively, 438.6 keV from $^{69}$Zn decay and the
line complex at 1105 - 1125 keV essentially from $^{65}$Zn and $^{69}$Ge decay.
These lines respresent a sample of very short (20.40 ms) to relatively long
decay half lifes (244 d) and can not be produced  in the solar flare. A
description of SPI background lines can be found in Weidenspointner et al.
\cite{weiden2}. The agreement in background to flare line ratio with the
respective effective data taking times was better than 1.5\% in all cases. We
adopted finally an uncertainty of 1.5\% for this normalization factor and added
it quadratically to the statistical errors. The resulting spectra of the three
different phases are shown in  fig.~\ref{fig3}.

In the spectrum of phase A, the 2.223 MeV line is very weak compared to phases B
and  C. The deexcitation lines can be hardly distinguished, while the
continuum  above 7 MeV seems comparatively stronger than in the two later
phases.  The line at 511  keV is also clearly visible but mainly of
instrumental origin, and is therefore not considered in the present analysis.
The same holds for the continuum below $\approx$600 keV, which is dominated by
Compton scattered photons of higher energy and photons originating from  pair
creation reactions in the passive structural material. Spectra from phases B
and C show a prominent narrow 2.223 MeV line, with a substantial Compton tail
extending down to about 600 keV as well as clearly  visible structures at 4.4
MeV and 6.1 MeV. They contain also the two $^{16}$O deexcitation lines around 7
MeV  (Gros et al. \cite{gros}) as well as the lines from $^{24}$Mg, $^{20}$Ne
and $^{28}$Si. These lines are statistically significant but not obviously
recognizable on the graphs.    

\section{Data analysis}

The instrumental response to a monoenergetic gamma-ray flux in the present
configuration, and especially in the several-MeV energy range is relatively
complex and cannot be easily deduced from the ground calibration campaigns.  A
detailed description of the instruments and the calibration can be found in
Vedrenne et al. \cite{vedrenne} and Atti\'e et al. \cite{attie}. The main
differences to the standard response function are the strongly reduced
full-energy peak efficiencies due to absorption and a continuum component from
forward Compton scattered photons in the passive and active material of the
satellite. Extensive Monte-Carlo simulations with the GEANT package and an
approximate model of the spacecraft showed that essentially only the material
close  to the line {Sun} - Ge matrix is responsible for this component
and in particular the matter sitting below the detector plane inside the BGO
shield.  These Compton scattered photons having lost only a small amount of
energy contribute considerably to the continuum just below the line, which is
of particular importance for the line shape calculations. 

We made therefore Monte-Carlo simulations with the code MGGPOD (Weidenspointner
et al. \cite{weidens}) using the detailed SPI mass model (Sturner et al.
\cite{sturner}) coupled to the TIMM model (Ferguson et al. \cite{ferguson}) for
the other parts of the spacecraft for gamma-ray energies corresponding to the
lines observed at 1.37, 1.63, 1.78, 2.223, 4.4, 6.1 and 7 MeV. Monoenergetic
beams of a projected area of 10098 cm$^2$ centered at the Ge-detector array
with 10$^7$ photons for each energy were generated in the simulations to obtain
the spectral response for the lines of interest. The results for the 
full-energy peak effective detector areas and the transmission probabilities  
are listed in table \ref{table1}. The transmission factors have been obtained
assuming that the intrinsic efficiency of the detector array in this
configuration equals the intrinsic efficiency of the Ge camera determined in
the ground calibration campaign at Bruy\`eres-le-ch\^{a}tel (Atti\'e et al.
\cite{attie}). Uncertainties of the effective area are estimated to be smaller
than 10\%.  Uncertainties concerning the ratios of effective areas for
different lines are estimated to be smaller than 5\%. 

\begin{table}  
\caption{Effective areas and transmission factors of the SPI gamma-ray
 camera for the observation conditions of the flare obtained by MGGPOD
 simulations with a detailed model of SPI and the spacecraft. 
 }
 \label{table1} 
\centering

\begin{tabular}  {llcc} 
\hline \hline
E$_{\gamma}$ (keV) & origin & eff. area (cm$^2$)  &
transmission (\%) \\
\hline
 \\ 
1369 & $^{24}$Mg$^{\star}_{1369}$ & 3.6 & 2.8  \\
1634 & $^{20}$Ne$^{\star}_{1634}$ & 4.6 & 4.0  \\
1779 & $^{28}$Si$^{\star}_{1779}$ & 5.0 & 4.5  \\
2223 &  $^1$H(n,$\gamma$)$^2$H & 5.4 & 5.7  \\
4438 & $^{12}$C$^{\star}_{4439}$ & 5.2 & 8.7  \\
6129 & $^{16}$O$^{\star}_{6130}$ & 4.1 & 9.3  \\
6916 & $^{16}$O$^{\star}_{6917}$ & 3.6 & 9.4  \\
7115 & $^{16}$O$^{\star}_{7117}$ & 3.6 & 9.4  \\
\hline
\end{tabular} 
\end{table}

\begin{figure} 
\resizebox{\hsize}{!}{\includegraphics{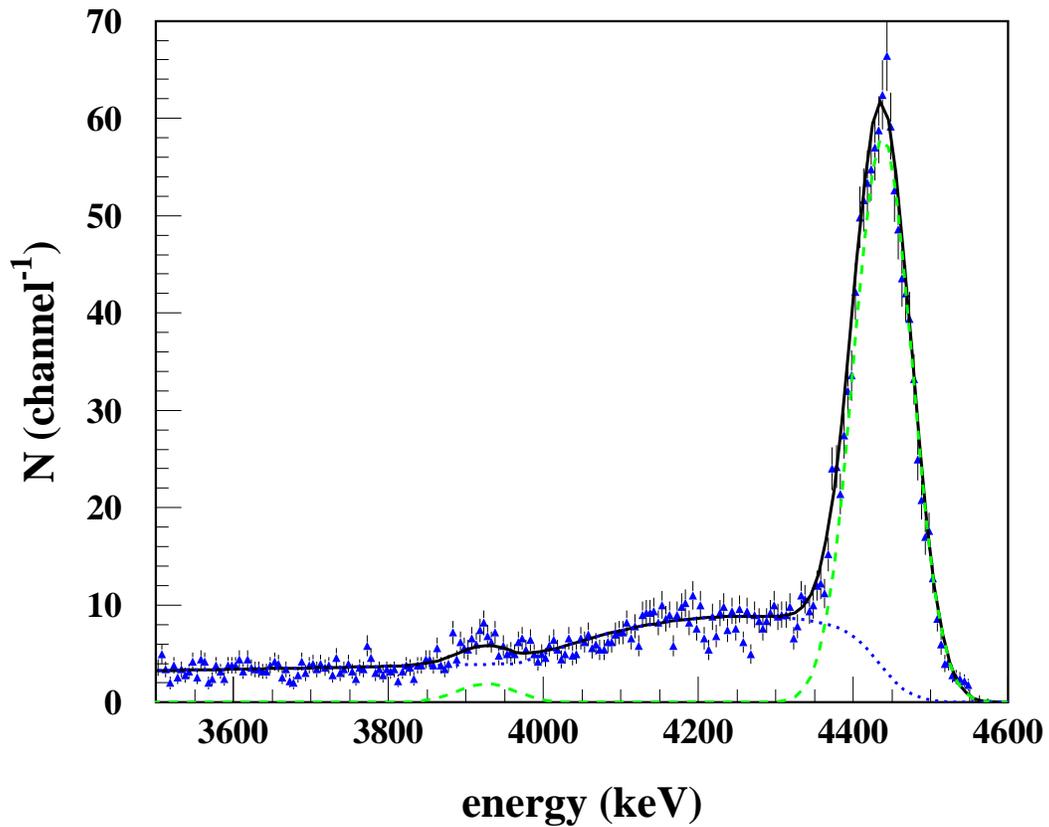}}

\caption{Symbols show the result of MGGPOD simulations for the response of SPI
for an incident flux of gamma-rays with  Gaussian  energy distribution centered
at 4438 keV and 88 keV FWHM for the particular observation conditions of the
Oct. 28 flare.  The different lines show the parameterization of the
instrumental response function used for the global fits to the observed spectra.
Dotted line: Compton component; dashed line: full-energy
and first-escape peak; full line: sum of the components.  }

\label{fig4}
\end{figure}

The spectra resulting from the simulations were then parameterized for the
purpose of a global fit of the observed lines and the continuum during the
different phases of the flare. The line components included a polynomial
function describing the Compton tail at low energies from 600 keV up to the 
first escape peak. No attempts were made to describe the spectra below that
energy region, for which the transmission gets anyway negligible. A second
polynomial function was used to describe the Compton component between the 
first escape peak and the full-energy peak, multiplied by a function 
simulating the steplike behaviour of the Compton component below the
full-energy peak. The first-escape and the full-energy peaks were taken as
Gaussians.  An example of the simulated spectrum for a gamma-ray line at 4438
keV with a FWHM of 88 keV and the described parameterization is shown in
fig.~\ref{fig4}.    

In a first analysis step, observed spectra of the different phases were fitted
with the parameterized response of the three strongest lines at 2.223, 4.4 and
6.1 MeV, leaving the amplitudes, widths and line centroids (except for the
fixed centroid of the 2223 keV line) as free parameters, and two  components
for the underlying continuum. The first component takes account of the
bremssstrahlung continuum from accelerated electrons.   A second continuum
component between 600 keV and 2.223 MeV  takes account of Compton scattering 
in the solar atmosphere of the 2.223 MeV gamma rays, because neutron capture on
hydrogen may occur in depths reaching 10 g/cm$^2$. The photon flux incident on
the satellite for this component was obained by simulating the transport of the
2.223 MeV gamma rays through the solar atmosphere with the GEANT package. We
used several depth distributions of the neutron capture as shown in Hua \&
Lingenfelter \cite{hua} and choose finally an averaged  photon spectrum
approximating fairly  well the different distributions. 

This spectrum as well as the component from electron bremsstrahlung for which
we choose an incoming power-law flux was then folded with the instrumental
response function obtained from the GEANT and MGGPOD simulations as described
above. We left the normalization factor for both continuum components  as free
parameters, limiting however the solar Compton scattering component to stay
below the photon flux  that was obtained from simulations using the deepest
neutron capture depth distribution  of Hua \& Lingenfelter \cite{hua} (centered
at about 10 g/cm$^2$).   This component, folded with instrumental response
function could be fairly well described by a second order  polynomial while
the bremsstrahlung continuum could be described by a power-law function
multiplied by an exponential above 2 MeV:

\begin{equation}
\frac{dN(E)}{dE} ~ \propto ~ E^{-s} ~ F_{exp}(E) 
\end{equation}

$F_{exp}(E)~ = ~ 1 ~~~~~~~~~~~~~~~~~~~~~~$ for E$<$2000 keV 

$F_{exp}(E) ~ = ~ e^{-0.0003 (E-2000)} ~~$ for E$\geq$2000 keV
\vspace{0.2 cm}

\begin{figure} 
\resizebox{\hsize}{!}{\includegraphics{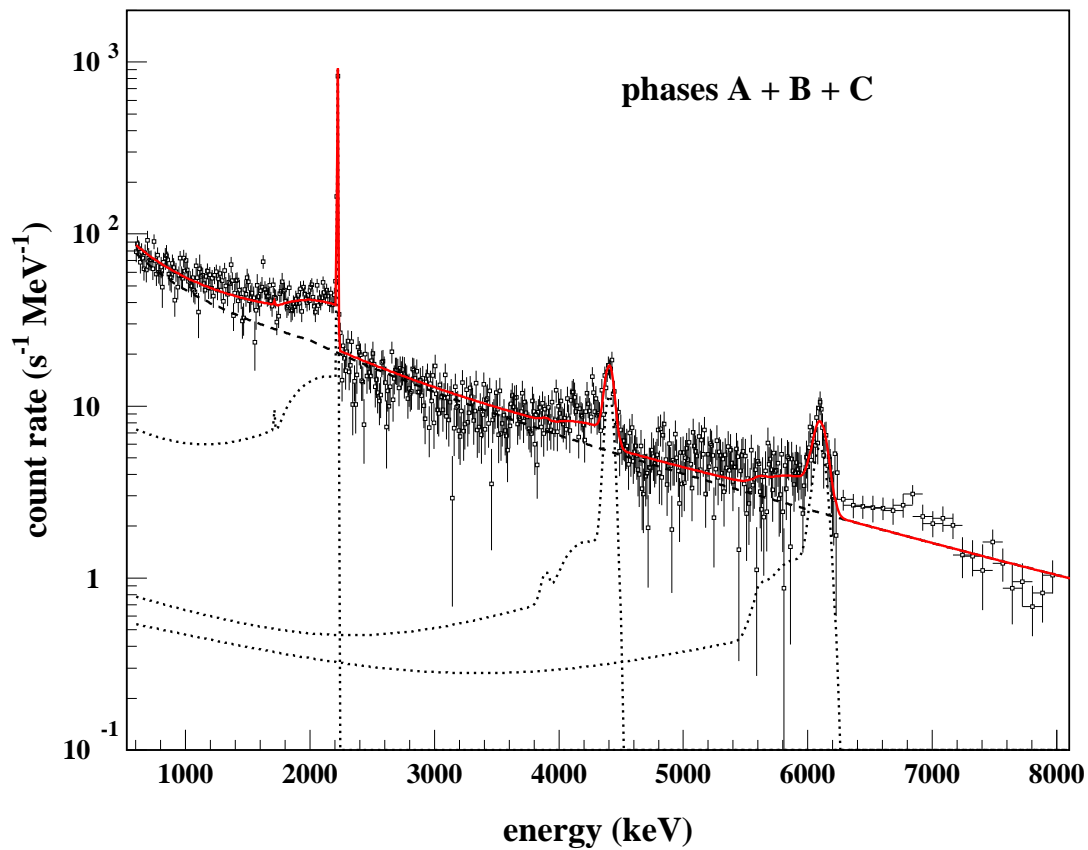}}
\caption{
Background subtracted spectrum of time intervals A-C (symbols) with the
result of the spectrum fit showing the sum of line and continuum components
(full line). The line complexes including full-energy and first escape peak
and the Compton tail are shown separately (dotted line). 
The excess around 7 MeV is due to the 6.9 and 7.1 MeV lines of $^{16}$O.The
dashed line  represents the continuum component.}
\label{fig5}
\end{figure}

Fits were performed for spectra of the three phases between 600 keV and 8 MeV,
excluding the energy bins 830-850, 1350-1370, 1610-1640, 1760-1780 keV and
6.8-7.2 MeV in phases B and C, where strong lines from $^{56}$Fe, $^{24}$Mg,
$^{20}$Ne, $^{28}$Si and $^{16}$O deexcitation are situated. Fig.~\ref{fig5}
shows the fit of the spectrum of the combined phases A, B and C, which include
the  main part of nuclear gamma-ray line emission of the flare. In the case of
the phase A spectrum we fitted first the continuum part with the 2.223 MeV line
complex, excluding the 4.4 and 6.1 MeV line ranges, and obtained then the
parameters for the latter two lines by a fit to the residuals. All other
spectra were fitted simultaneously with the continuum and the three lines.  The
spectra are fairly well fit throughout the whole region with these components.
Results of the spectrum fits are presented in table \ref{table2}. 

\begin{table*} 
 \caption{
Results of the spectrum fits  for the three strongest lines and the underlying
continuum. Line centroids and $\sigma$ are in keV, rates are dead-time corrected
counts/s and line fluences are expressed in photons/cm$^2$. S is the power-law
index of the observed photon spectrum. Error bars result
from the fit with an additional uncertainty of 10\% for the fluences, due to an
uncertainty on  the effective area, added quadratically to the error bar of the
fit.}
\label{table2} 
\centering

\begin{tabular}  {clccccc} 
\hline \hline
 & phase  & A & B & C & A-C & \\
\hline
\\
continuum & s  & 0.87$^{+0.02}_{-0.02}$ & 0.98$^{+0.02}_{-0.02}$  &
1.11$^{+0.03}_{-0.03}$ & 0.97$^{+0.02}_{-0.02}$  & \\
\\
(0.6-8 MeV) & rate & 174$^{+7}_{-2}$ & 142$^{+3}_{-3}$ & 59$^{+2}_{-2}$ &
95$^{+2}_{-2}$ & \\
\\
\\
$^1$H(n,$\gamma$) & rate & 2.19$^{+0.39}_{-0.42}$ & 15.2$^{+0.5}_{-0.5}$
 & 8.66$^{+0.24}_{-0.24}$ & 9.65$^{+0.21}_{-0.21}$ & \\ 
 \\
 & fluence & 30$^{+6}_{-7}$ &  549$^{+58}_{-58}$ &  721$^{+75}_{-75}$ &
 1286$^{+132}_{-132}$ & \\    
\\
\\
$^{12}$C$^{\star}$ & centroid & 4400$^{+10}_{-11}$ &  4398$^{+5}_{-6}$  &
4415$^{+6}_{-6}$ & 4407$^{+4}_{-4}$ & \\
\\
 & $\sigma$ & 28$^{+12}_{-6}$ & 40$^{+8}_{-6}$  & 40$^{+8}_{-6}$  &
37$^{+5}_{-4}$  & \\
\\
 & rate & 0.80$^{+0.42}_{-0.35}$ &  1.86$^{+0.28}_{-0.27}$ &
 0.72$^{+0.12}_{-0.12}$ & 1.01$^{+0.12}_{-0.11}$ & \\
 \\
 & fluence &  12$^{+6}_{-5}$ &  70$^{+13}_{-12}$ &  62$^{+12}_{-12}$ & 
 140$^{+21}_{-21}$  & \\
 \\
 \\
$^{16}$O$^{\star}$ & centroid & -  & 6093$^{+9}_{-8}$  & 
6099$^{+7}_{-6}$  & 6102$^{+7}_{-7}$ &   \\
\\
 & $\sigma$ & - & 59$^{+10}_{-8}$  & 31$^{+9}_{-7}$  & 
57$^{+8}_{-7}$ &   \\
\\
 & rate &  -    & 1.50$^{+0.20}_{-0.20}$  &
 0.35$^{+0.08}_{-0.08}$ & 0.74$^{+0.09}_{-0.09}$  &   \\
 \\
 & fluence &  -  &  71$^{+12}_{-12}$  &  38$^{+9}_{-9}$ & 
 130$^{+21}_{-21}$  &   \\
 \\ 
$\chi^2_{d.o.f}$ &  & 1.06 & 1.25 & 1.08 & 1.24 & \\
\hline
\end{tabular}
\end{table*}
 
With the exception of the  $^{16}$O line in phase A statistically significant
excess for the three studied gamma-ray lines could be observed during all
phases. Line fluences for phases A-C comprise more than 95\% of the total
fluence for the two deexcitation lines and of about 90\% for the
neutron-capture line. With a total 2.223 MeV line fluence of about 1400
photons/cm$^2$, the October 28, 2003 flare is probably the one with the highest
ever observed fluence for the neutron-capture line, exceeding the highest one
observed by SMM on October 24, 1989 (Verstrand et al. \cite{verstrand}) by a
factor of two and the one of the very  intense June 4, 1991 solar flare
observed by OSSE   (Murphy et al. \cite{mur97}) by about 40\%. 

There is a clear evolution of increasing
spectral index s for the continuum from phases A to C and a rapidly increasing
line-to-continuum ratio in the count rates between A and B, staying
approximately stable afterwards. There is also evidence for a larger redshift
of the $^{12}$C line - and to a lesser extent of the $^{16}$O line -  during
phase B compared to phase C and there is possibly a change in the 4.4 MeV to
6.1 MeV line flux ratio from phase B to C. We concentrate in the following on
these two phases with strong nuclear line emission for a detailed line shape
analysis.   

\section{Line shape analysis}

We calculated the profile of the $^{12}$C line as described in Kiener et al.
\cite{kiener}, where the 4.438 MeV gamma-ray production
by accelerated protons and $\alpha$-particles interacting with $^{12}$C and
$^{16}$O was based on extensive experimental data on total and differential
cross sections and on measured line shapes. A similar approach was chosen for
the 6.129 MeV gamma ray of $^{16}$O, produced by inelastic scattering
off $^{16}$O and spallation of $^{20}$Ne. The required nuclear reaction
parameters were deduced from available cross section data, nuclear structure
information and optical model calculations. We included also the 6.175 MeV line
of  $^{15}$O, produced mainly by spallation of  $^{16}$O  which could
contribute to the high-energy tail of the 6.1 MeV line profile. 

Due to uncertainties of the nuclear reaction models, the predicted line
profiles F$_{calc}$(E) have to  be taken with some caution. A comparison of
calculated and measured line shapes in an accelerator experiment for the 4.4
MeV line at different angles with respect to incoming proton beam can be found
in Kiener et al. \cite{kiener}. We attribute therefore an uncertainty to the
calculated line by assigning errors to the calculated profile
$\Delta$F$_{calc}$(E)~=~ k ~$\times$ ~F$_{calc}$(E) .  A
conservative estimate for k on the calculated line shape of the 4.4
MeV line is 0.2, whose calculations could be largely based on measured line
shapes in the important energy range and 0.4 for the 6.1 MeV line, where no
measured line shapes are available. Practically, we added these uncertainties 
quadratically to the error bars of the data for the line shape analyses.

All line shape calculations were based on a thick-target interaction model and
were done with two different abundance sets of C, O and  Ne. The first set are
the chromospheric abundances of  Reames \cite{reames}. These abundances are 
compatible with the observed fluence ratio between the carbon and the oxygen
line $F_{4.4}/F_{6.1}$  in phase C; in a second set, we reduced the abundance
of carbon by 50\% with respect to oxygen, which fits better the line ratios of
phase B and of the combined phases B-C. The same abundances relative to
hydrogen  were also used for the accelerated heavy ions producing broad lines
by interacting with ambient hydrogen and helium nuclei with He/H = 0.1. These
components are however typically a factor of five to ten smaller than the
Compton component of the narrow lines in the energy region used for the  fits.
We therefore did not explore different accelerated heavy-ion to proton ratios.
$\alpha$/p was left as a free parameter. 

We added then to the calculated line profiles the Compton tails from the MGGPOD
simulations and the continuum component from the global  fit to the spectrum of
the respective time  interval. Comparison with the observed line profiles was
done in the energy ranges 4280-4550 keV and 5955-6225 keV for the  $^{12}$C and
the $^{16}$O line, respectively. 
For the energy distribution of accelerated particles we used the same power-law
spectrum  in energy per nucleon for each species with an exponential cutoff at
200 MeV per nucleon. The spectral index s was deduced from the 2.223 MeV to the
4.4 and 6.1 MeV line fluence ratios to be in the range 3 $\leq$ s $\leq$ 4
(Tatischeff et al. \cite{tatische}). Share et al.\cite{share04} find s
$\approx$ 3  for the decay phase of the same flare . We made calculations with
s = 3.0, 3.5 and 4.0. Three different particle angular distributions were
used:

\noindent
(1) A downward-directed (DW) distribution, where the angular
distribution  of energetic particles around the flare normal is given by: 

\begin{equation}
\frac{dN}{d\Omega}~ \propto ~ e^{-\Theta/\Delta \Theta}
 \end{equation}

with the flare normal taken to be perpendicular to the {Sun}'s surface 
and pointing towards the center of the {Sun}. 

\noindent
(2) A downward-isotropic (DI) distribution with:

\begin{equation}
\frac{dN}{d\Omega}~ = ~ const. ~~~  \Theta \leq 90^{\circ} \newline
 \frac{dN}{d\Omega}~ = ~ 0 ~~~  \Theta > 90^{\circ} 
 \end{equation}

\begin{figure} 
\resizebox{\hsize}{!}{\includegraphics{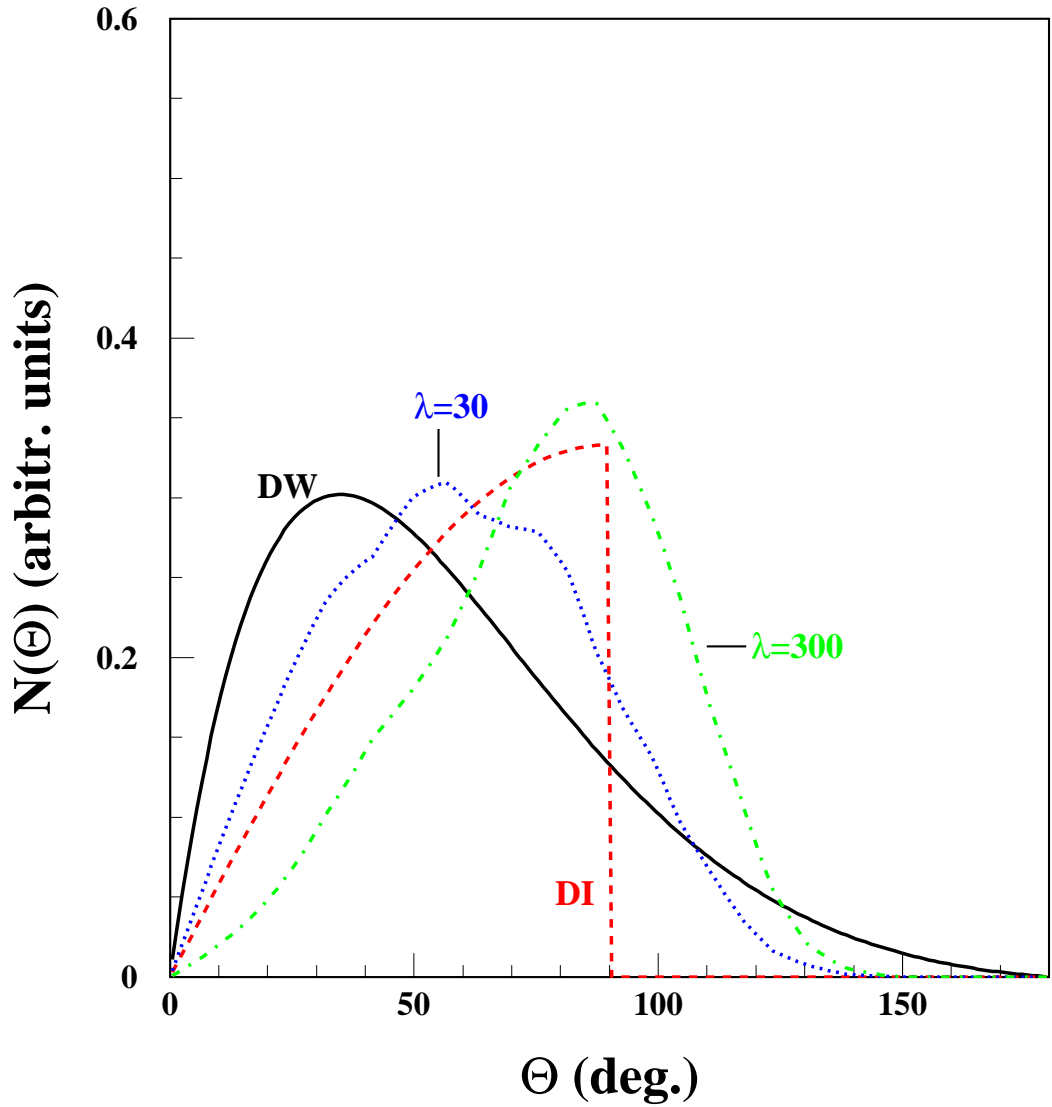}}
\caption{Relative number of interacting particles as a function of $\Theta$ 
used for the
line shape calculations. $\Theta$ is the angle with respect to the flare axis,
defined as the normal to the sun's surface at the flare location. The DW
distribution is for $\Delta \Theta$ = 40$^{\circ}$}
\label{fig6}
\end{figure}

\noindent 
(3) Pitch-angle scattering (PAS) distributions of Murphy et al.
\cite{mur90} with  mean-free path parameters $\lambda$ = 30 and 300. They are based
on a model of particle transport in coronal magnetic loops with constant field
and converging field lines in the solar chromosphere and photosphere. Particles
are injected isotropically into the loop and transported in the magnetic field
taking account of pitch-angle scattering on MHD turbulence. $\lambda$ is
defined in this model as the scattering mean free path divided by the
half-length of the coronal segment of the magnetic loop. The angular
distributions of the interacting particles producing gamma rays are governed by
the competition  between magnetic mirroring and pitch-angle scattering.
Magnetic mirroring is dominant for $\lambda$ = 300 and the distribution 
approaches a fan beam  parallel to the solar surface, while strong pitch-angle
scattering $\lambda$ = 30 gives downward directed distributions. The different
angular distributions are plotted in fig.~\ref{fig6}.

For the location of the flare, coordinates given by GOES for the start of the
flare give a heliocentric angle $\Theta_h$  = 30$^{\circ}$;  RHESSI could
resolve  two emitting regions centered at $\Theta_h$  = 22$^{\circ}$ and
$\Theta_h$  = 25$^{\circ}$ in the hard-X ray and 2.223 MeV gamma-ray line
contours during the decline phase of the flare (Hurford et al. \cite{hurford}).
The exact position of the flare having practically no influence on the
calculated line shapes we restricted calculations to one heliocentric angle.
All following calculations were done with  $\Theta_h$  = 25$^{\circ}$ varying
systematically  $\alpha$/p and $\Delta \Theta$ in the case of DW distributions.
A $\chi^2$ was obtained for each parameter  set with a normalization factor for
the calculated line shape adjusted to minimize   $\chi^2$  with the MINUIT
package of CERN.  Estimated uncertainties of the theoretical
line shapes as discussed above and of 10\% for the
Compton components were added quadratically to the error bars of the data. 

We explored first the 4.4 and 6.1 MeV lines separately to get a new
determination of the line fluences and test the validity of the results
obtained with a Gaussian distribution.  Best fits for the lines in time
intervals B, C and the combined B-C and A-C intervals were mostly obtained with
relatively narrow DW distributions of particles. However,  DI and
PAS(${\lambda=30}$) distributions in many cases gave also acceptable fits.
Best-fit values of count rates and fluences are given in table \ref{table3}. 
Count rates obtained with this method agree with the ones obtained by global
fits of the spectra using Gaussian line shapes but are slightly  more accurate
despite the fact that the given error bars encompass all 1-sigma uncertainties
of the different fits exploring a vast parameter space in spectral index,
$\alpha$/p and particle-angular distributions. This analysis confirms that
there is strong evidence for a change in the line fluence ratio, the ratio of
count rates  evolving from R$_{4.4}$/R$_{6.1}$ =  1.25$^{+0.13}_{-0.13}$ in
phase B to R$_{4.4}$/R$_{6.1}$ =  1.66$^{+0.25}_{-0.25}$ in phase C.    

\begin{table}  
\caption{
Results of separate fits to the $^{12}$C-line at 4.4 MeV and the $^{16}$O-line
line at 6.1 MeV assuming a thick-target interaction model and three different
energetic particle angular distributions around the flare axis. 
Rates and  fluences are expressed as in table I. Error bars on count
rates encompass the individual 1-sigma uncertainties of the different fits; a
10\% uncertainty has been quadratically added to the error bars of line
fluences.}
\label{table3} 
\centering

\begin{tabular}  {clcccc} \hline \hline
& phase  & B & C & B-C & A-C  \\
\hline
\\
$^{12}$C$^{\star}$ & $\alpha$/p & 0.19 & 0.04 & 0.10 & 0.09  \\

& $\Delta \Theta$ & 25$^{\circ}$ & 40$^{\circ}$ & 35$^{\circ}$  &
30$^{\circ}$  \\
 \\
 & rate & 2.03$^{+0.15}_{-0.17}$ &  0.68$^{+0.09}_{-0.06}$ &
 1.09$^{+0.09}_{-0.07}$ & 1.02$^{+0.10}_{-0.05}$  \\
 \\
 & fluence & 76$^{+9}_{-10}$  &  58$^{+9}_{-8}$ &  127$^{+16}_{-15}$ &
 141$^{+20}_{-16}$  \\
 \\
 \\
$^{16}$O$^{\star}$ & $\alpha$/p &  0.20 &  0.02 & 0.08 & 0.11   \\
&  $\Delta \Theta$ & 60$^{\circ}$ & 40$^{\circ}$ & 55$^{\circ}$ &
70$^{\circ}$  \\
   \\
 & rate & 1.62$^{+0.12}_{-0.09} $  & 0.40$^{+0.06}_{-0.03}$  &
 0.77$^{+0.06}_{-0.04}$ &  0.74$^{+0.06}_{-0.03}$    \\
 \\
 & fluence & 77$^{+10}_{-9}$ &  44$^{+8}_{-6}$  &  121$^{+15}_{-14}$ &
 130$^{+17}_{-14}$    \\
 \\ 
\\
 $^{24}$Mg$^{\star}$ & fluence & 45$^{+15}_{-14}$ &  28$^{+18}_{-18}$  & 
 75$^{+24}_{-24}$ & 74$^{+28}_{-24}$    \\
 \\ 
\\
$^{20}$Ne$^{\star}$ & fluence & 34$^{+13}_{-11}$ &  30$^{+14}_{-13}$  & 
 63$^{+24}_{-19}$ & 71$^{+28}_{-20}$    \\
 \\ 
\\
$^{28}$Si$^{\star}$ & fluence & 21$^{+10}_{-9}$ &  20$^{+11}_{-11}$  & 
 43$^{+17}_{-14}$ & 50$^{+19}_{-16}$    \\
\\
 \hline

\end{tabular} 
\end{table}

\begin{figure} 
\resizebox{\hsize}{!}{\includegraphics{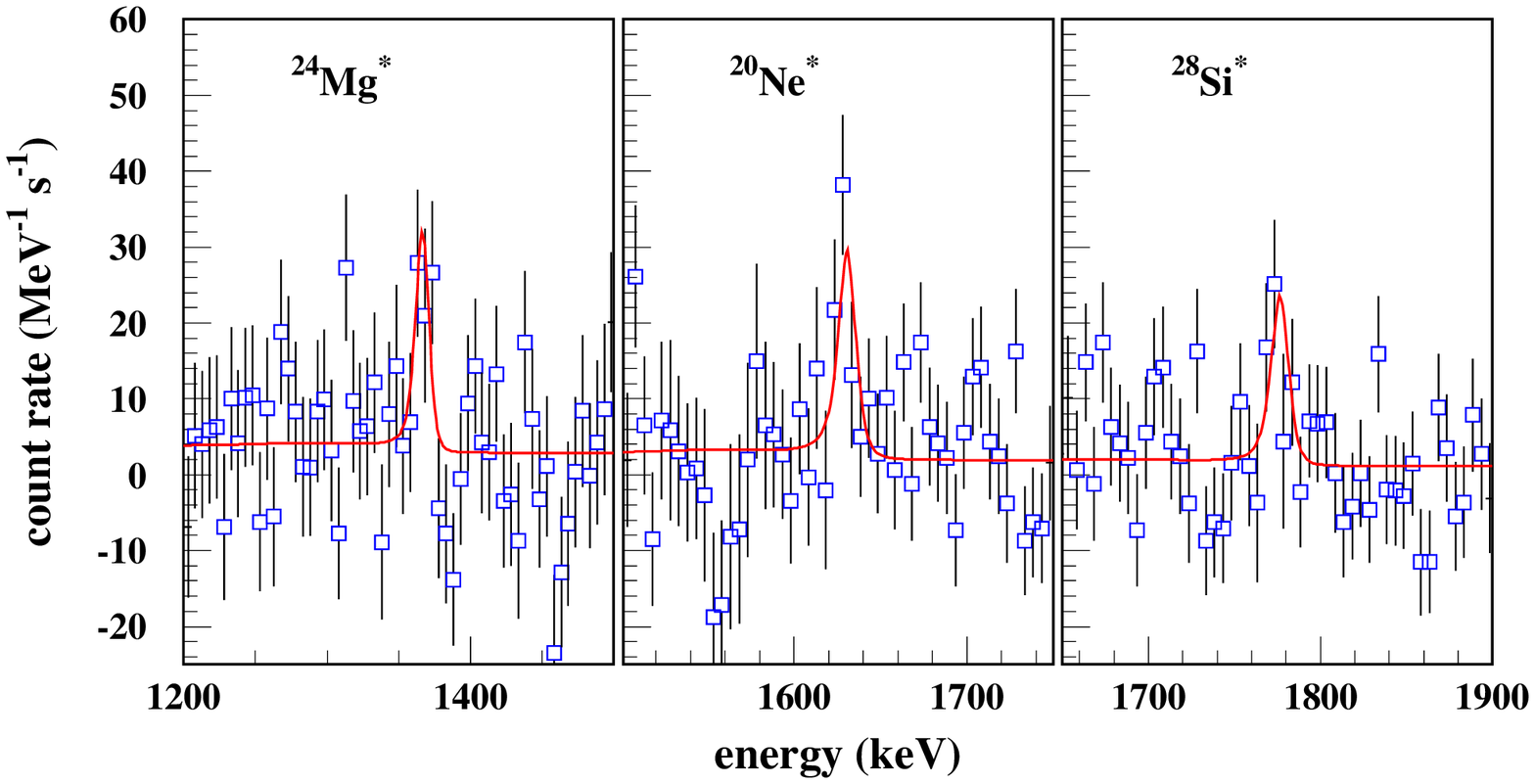}}
\caption{Calculated line shapes of the 1.37, 1.63 and 1.78 MeV lines for 
DW distributions with best fit parameters from combined fits to the 4.4 and 6.1
MeV lines. Data points represent the residual data of phases B-C after
subtraction of the continuum components and the Compton tails of the lines at
2.223, 4.4 and 6.1 MeV.}
\label{fig7}
\end{figure}

A similar approach was chosen to obtain the line fluences of the 1.37, 1.63 and
1.78 MeV gamma rays from $^{24}$Mg, $^{20}$Ne and $^{28}$Si, respectively. We
used the residual spectra  for the fits. These spectra were obtained
by subtracting the bremsstrahlung and solar Compton continuum and the Compton
components of the three strongest lines at 2.223, 4.4 and 6.1 MeV obtained from
the global fits (table \ref{table2}). The energy regions around the
respective lines were then fitted with a constant to take account of an
eventual residual background and the calculated line shape which was folded with
the instrumental response as in the case of the three strongest lines. For the
line shape calculations we used available cross section data and proceeded 
otherwise as described in Ramaty et al. \cite{ramaty}. 
Line shapes of the three lines with best fit DW distributions are shown in fig.
~\ref{fig7}.

Results for the count rates and line fluences are quoted in table \ref{table3}.
The uncertainties comprise all 1-sigma error bars from fits with three
different energetic particle spectral indices, DW, DI and PAS angular
distributions and with the best fit parameters  $\alpha$/p and $\Delta \Theta$
that were obtained from combined fits to the 4.4 and 6.1 MeV lines as described
below. Relative line fluences of the five lines are
compatible with the ones observed by RHESSI for the July 23, 2002 flare (Smith
et al. \cite{smith}) and with the exception of the $^{20}$Ne line also
compatible with the results from various flares observed by CGRO/OSSE and SMM
(Murphy et al. \cite{mur97}). It is interesting to note that the two satellites
equipped with high-resolution Ge detectors give consistently a factor of about
two smaller relative $^{20}$Ne fluxes than the scintillation instruments.

Determinations of $\alpha$/p and the angular
distributions were done by fitting simultaneously  the 4.4 and 6.1 MeV lines.
The same power-law energy spectra and the two different carbon abundances were
used as in the case of individual fits to the lines. We like to stress that the
line fluence ratio is not a free parameter in these fits but calculated with
the thick-target interaction model used in the line shape calculations. This
includes also the effect of the gamma-ray angular distribution with respect to
the flare axis, which is different for 4.4 and 6.1 MeV gamma-ray emission,
especially for very asymmetric energetic particle angular  distributions as
e.g. the narrow DW distributions. As expected for data of time intervals B, B-C
and A-C, fits with the 50\% reduced carbon-to-oxygen abundance ratio C/O gave
systematically smaller $\chi^2$ than fits with nominal abundance, the
difference being  $\Delta \chi^2 ~ \approx ~ $ 2-3. Surprisingly, combined fits
of both lines had also slightly smaller $\chi^2$  ($\Delta \chi^2 ~ \approx ~ $
0.5-1.5) with reduced C/O for phase C, where the line fluence
ratio does not necessarily favor a smaller carbon  abundance. The following
results were obtained with reduced C/O.

\begin{table*} 
 \caption{
Best fit values of angular distribution width parameter $\Delta \Theta$ for 
DW flares and of $\alpha$/p for DW and PAS($\lambda=30$) flares 
for combined fits of the $^{12}$C and $^{16}$O lines.
Spectral index of accelerated particles s=4.0; values in parenthesis are for  
s=3.0. The values of $\alpha$/p and $\Delta \Theta$ 
for s=3.5 are generally between those of s=3.0 and s=4.0}
 \label{table4} 
\centering

\begin{tabular}  {lccccc} 
\hline \hline
flare geometry & \multicolumn{3}{c}{DW} &  \multicolumn{2}{c}{PAS} \\
 & $\alpha$/p & $\Delta \Theta$ & $\chi^2$ & $\alpha$/p
 & $\chi^2$ \\ 
\hline \
\\ \\ 
phase B & 0.11 (0.19) & 35$^{\circ}$ (30$^{\circ}) $ &  48.5 (48.8) & 0.18 
(0.35)  & 49.9 (52.1) \\

phase C & 0.06 (0.00) & 40$^{\circ}$ (40$^{\circ}$) & 25.1 (26.8)  &   0.10
(0.03) & 26.0 (27.2) \\

phase B-C & 0.07 (0.11) & 40$^{\circ}$ (35$^{\circ}$) & 27.8 (28.7) & 0.11 
(0.19) & 29.9 (31.3)   \\
\hline
\end{tabular}

\end{table*}

\begin{figure} 
\resizebox{\hsize}{!}{\includegraphics{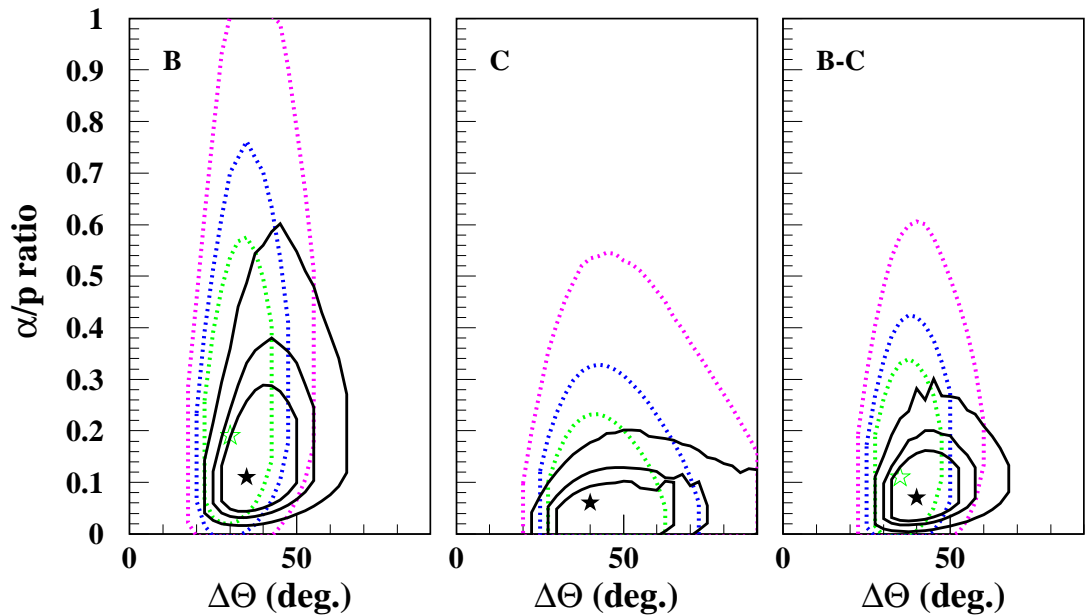}}
\caption{
Contour lines in the $\alpha$/p - $\Delta \Theta$ parameter space of 
downward-directed particle distributions for simultaneous
fits  to
the 4.4 and 6.1 MeV lines in flare phases B, C and B-C. Plotted are
50\%, 70\% and 90\% confidence levels for fits using
the reduced C/O and power-law particle spectra with 
s = 3.0 (dotted gray lines) and s = 4.0 (continuous black lines). $\chi^2$
minima are indicated by open (s = 3.0) and filled symbols (s = 4.0). The open
symbol for phase C at $\alpha$/p = 0.00 $\Delta \Theta$ =
30$^{\circ}$ is not plotted.}
\label{fig8}
\end{figure}

Best fits for all time intervals were obtained with a narrow DW distribution of
the accelerated particles with $\Delta \Theta \approx$ 40$^{\circ}$ and
$\alpha$/p  $\approx$ 0.1. The 50\%, 70\% and 90\% confidence limit contours
for these two parameters are shown in fig.~\ref{fig8}. While the angular
distribution of the energetic particles in phases B and C seems unchanged, an
evolution of $\alpha$/p is possible with best fit values of 0.11 - 0.19 in
phase B and a 90\% upper limit of $\approx$ 1.1 to very low best fit values of
0.00 - 0.06 in phase C,  with a 90\% upper limit at $\approx$ 0.6.
These lower $\alpha$/p fit values result from  smaller redshifts of the 4.4 and
6.1 MeV line and a smaller width of the 6.1 MeV line in phase C. These facts
could also be explained by  differences in flare geometry or an evolution of
the  power-law index, however, a change in $\alpha$/p would explain also
naturally the different line ratios. For example, reducing $\alpha$/p from 0.2
to 0.00 increases the C/O line ratio by 25\%, which is consistent with the
line-ratio evolution between phase B and C. This can hardly be achieved by
varying power-law index or flare geometry, and the only other explanation would
be a surprising abundance change of C or O in the interaction region.

\begin{figure} 
\resizebox{\hsize}{!}{\includegraphics{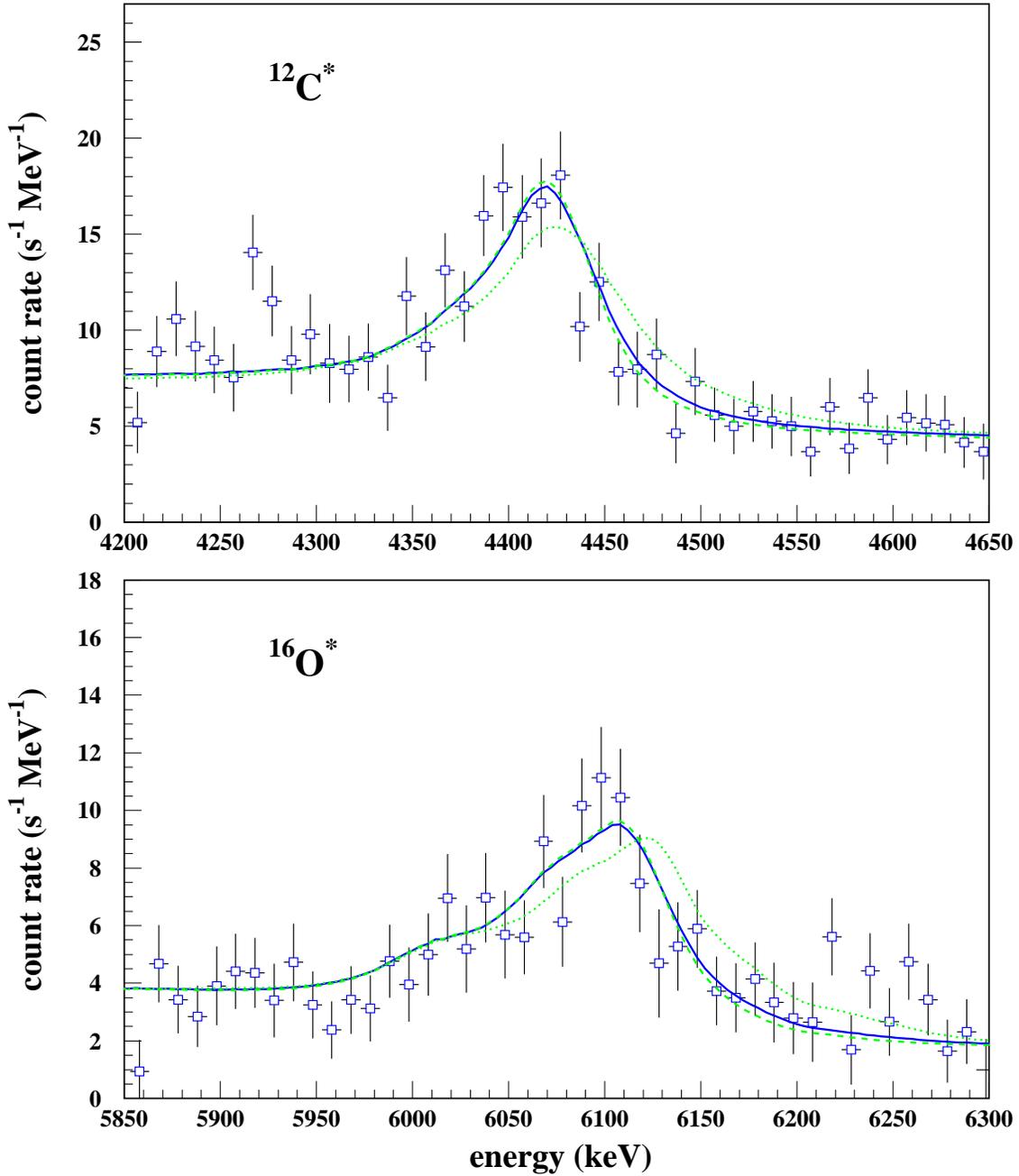}}
\caption{
Calculated line shapes for best fit parameters of DI (dashed
line), PAS($\lambda=30$) (full line) and PAS($\lambda=300$) (dotted line) 
energetic particle distributions in
comparison with observational data of time interval B-C. The line shapes of
PAS($\lambda=30$) and DI distributions are practically
indistinguishable.}
\label{fig9}
\end{figure}

\begin{figure*}  
\centering 
\includegraphics[width=17 cm]{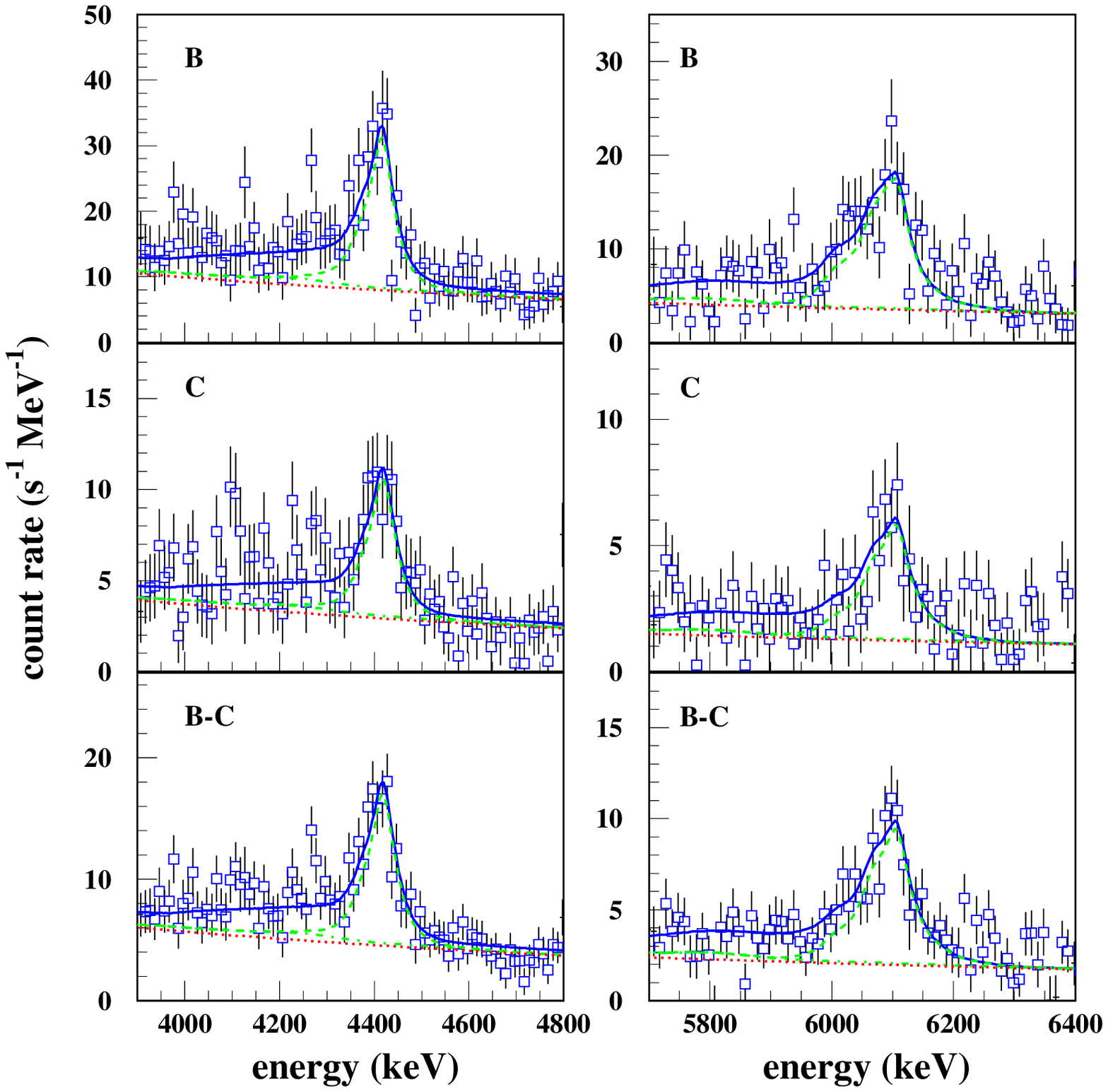} 
\caption{ Calculated line shapes  of the $^{12}$C
line at 4.4 MeV (left) and the
 $^{16}$O line at 6.1 MeV (right) for best fit parameters of downward-directed
flares in comparison with observational data for time intervals B, C and B-C.
Dotted line: continuum component, dashed-dotted line: continuum plus broad line
component, dashed line: continuum plus narrow and broad line component, full
line: sum of line, continuum and Compton component.} 
\label{fig10}  
\end{figure*}

We investigated also the line shapes resulting from DI and PAS distributions.
With the pitch-angle scattering distributions of the model of Murphy et al.
(1990), the distribution with $\lambda$ = 300 could be excluded at the 99.7\%
level in phases B, B-C and A-C and at the 95.4\% level in phase C. Fits with  
$\lambda$ = 30 described the lines not as good as the best fit DW 
distributions but stayed generally inside the 1-sigma confidence level with
slightly  higher $\alpha$/p.  DI distributions gave similar $\chi^2$ values and
$\alpha$/p  as PAS(${\lambda=30}$) ones. Examples of calculated line shapes
representing best fit DW distributions are presented in fig.~\ref{fig10};  line
shapes for  best fit PAS and DI  distributions are shown in fig.~\ref{fig9}.
One can see in fig.~\ref{fig10} that observed line shapes in the fit region
depend practically only on the narrow component, the broad component from
accelerated heavy-ion interactions being even significantly smaller than the
Compton tail. 

All line shapes
are  quite nicely reproduced by the calculated ones, the reduced $\chi^2$
values being about 0.5 for phases C and B-C and about 1.0 for phase B, except
for PAS($\lambda=300$) distributions which provide bad fits to the data. Such
low $\chi^2$ values can be explained by our estimated uncertainties of
theoretical shapes which were added to the error bars of the data. The
theoretical uncertainty is  particularly important for the 6.1
MeV line, in fact comparable with its statistical uncertainty.  
This illustrates that still better  constraints could be obtained
from observations with the new generation gamma-ray satellites by improved
knowledge of some important nuclear reactions leading to strong gamma-ray
emission in solar flares.

\section{Conclusion}

The very intense gamma-ray flare of October 28, 2003 was observed by the
gamma-ray spectrometer SPI onboard INTEGRAL in the energy range from 600 keV to
8 MeV and by the Anti-Coincidence Shield of SPI above 150 keV. Time profiles
show three different phases of gamma-ray emission. In the first minute, an
emission peak was observed consisting mainly of continuum emission from
electron bremsstrahlung with very little nuclear line emission. In a second
phase which lasted $\approx$3.5 minutes, high flux levels containing  several
peaks with  strong nuclear line and continuum emission could be observed,
before a smooth decay phase of a about ten minutes of both emission components
terminates the gamma-ray flare. 

A power-law bremsstrahlung component and three prominent narrow nuclear lines
at 2.223 MeV from neutron capture on hydrogen, at 4.4 MeV from $^{12}$C
deexcitation and at 6.1 MeV line from $^{16}$O deexcitation were observed
during the three phases. There is clear evidence of a change in continuum
power-law index and the continuum-to-line ratio during the flare. More
surprisingly, the  flux ratio of the 4.4 MeV to 6.1 MeV lines shows a
significant change between the second and third phase of the flare. This line
ratio evolution could be confirmed by detailed line shape analyses of both
nuclear deexcitation lines based on thick-target interaction model of the
accelerated particles with the solar atmosphere and a set of recent nuclear
data and reaction calculations. The overall line ratio, in particular the line
ratio of the second phase, points to a significantly reduced carbon
abundance with respect to oxygen in the composition of the target solar
atmosphere.

Best fits of both lines resulting from a systematic scan of $\alpha$/p and the
width of the particle angular distribution indicated equally an evolution
between the second and third flare phase. While both phases clearly favor a
narrow downward-directed particle beam perpendicular to the solar surface,
$\alpha$/p  could have been higher in the second than in the third phase. Such
a drop of the energetic $\alpha$-particle content would also naturally explain
the observed change in the 4.4 MeV to 6.1 MeV line ratio between both phases.
We conclude that the simplest and most probable explanation is an evolution of
$\alpha$/p during the second and third phase of the flare. 

The observation of solar flares with high-resolution gamma-ray spectrometers,
like the ones of RHESSI and INTEGRAL, both launched in 2002 opened up new
possibilities of detailed line shape analyses to constrain  properties of the
accelerated particle distributions. In this paper, we could obtain tight
constraints on the particle angular distributions and interesting evidence for
temporal evolution of the alpha-to-proton ratio. These studies will now be
applied to other recently observed  flares.

\section{Acknowledgements}

Based on observations with INTEGRAL, an
ESA project with instruments and science data centre funded by ESA
member states (especially the PI countries: Denmark, France, Germany,
Italy, Switzerland, Spain), Czech Republic and Poland, and with the
participation of Russia and the USA.
We thank A. Bykov, P.I. of the INTEGRAL observation during  the flare,
who gave us rapid access to the data.


\begin{thebibliography}{99}

\bibitem{attie}
Atti\'e, D., Cordier B., Gros, M., et al. 2003, Astronomy and Astrophysics
411, L71

\bibitem{ferguson}
Ferguson C., Barlow, E.J., Bird, A.J., et al. 2003, 
Astronomy and Astrophysics 411, L19

\bibitem{gros}
Gros, M., Tatischeff, V., Kiener, et al 2004, in  
Proc. 5$^{th}$ INTEGRAL Workshop, the INTEGRAL Universe, Munich, Germany,
16-20 February 2004, ed. V. Sch\"{o}nfelder, G. Lichti \& C. Winkler, 
ESA SP-552, 669 

\bibitem{hua}
Hua, X.-M., \& Lingenfelter, R.E. 1987, Solar Physics 107, 351

\bibitem{hurford}
Hurford, G.J., Krucker, S., Lin, R.P., Schwartz, R.A., Share, G.H., \& Smith,
D.M. 2005, Astrophysical Journal Letters, in preparation

\bibitem{kiener}
Kiener, J., de S\'er\'eville, N., \& Tatischeff, V. 2001, Physical Review C 64,
025803

\bibitem{mur88}
Murphy, R.J., Kozlovsky, B., \& Ramaty, R. 1988, Astrophysical Journal 331, 1029

\bibitem{mur90}
Murphy, R.J., Hua, X.-M., Kozlovsky, B., \& Ramaty, R. 1990, Astrophysical 
Journal 351, 299

\bibitem{mur97}
Murphy, R.J., Share, G.H., Grove, J.E., Johnson, W.N., Kinzer R.L., Kurfess,
J.D., Strickman, M.S., \& Jung G.V. 1997 Astrophysical Journal 490, 883 

\bibitem{ramaty}
Ramaty, R., Kozlovsky, B., \& Lingenfelter, R.E. 1979, Astrophysical Journal 
Suppl. Ser., 40, 487

\bibitem{reames}
Reames, D.V 1999, Space. Sci. Rev., 90, 413

\bibitem{share97}
Share, G.H., \& Murphy, R.J. 1997, Astrophysical Journal 485, 409

\bibitem{share02}
Share, G.H., Murphy, R.J., Kiener, J., \& de S\'er\'eville, N. 2002, 
Astrophysical Journal 573, 464 

\bibitem{share03}
Share, G.H., Murphy, R.J., Skibo, J.G., Smith, D.M., Hudson, H.S., Lin, R.P.,
Shih, A.Y., Dennis, B.D., Schwartz, R.A., \& Kozlovsky, B. 2003, Astrophysical 
Journal  595, L85

\bibitem{share04}
Share, G.H., Murphy, R.J., Smith, D.M., Schwartz, R.A., \& Lin, R.P. 2004, 
Astrophysical Journal 615, L169

\bibitem{smith}
Smith, D.M., Share, G.H., Murphy, R.J., Schwartz, R.A., Shih, A.Y., \& Lin,
R.P. 2003, Astrophysical Journal 595, L81

\bibitem{sturner}
Sturner, S.J., Shrader, C.R., Weidenspointner, G., Teegarden B.J., et al. 
2003, Astronomy and Astrophysics 411, L81

\bibitem{tatische}
Tatischeff, V., Kiener, J., \& Gros, M. 2004, in 
Proceedings 5$^{th}$ Rencontres du Vietnam, New Views on the Universe, Hanoi,
Vietnam 

\bibitem{vedrenne}
Vedrenne, G., Roques, J.-P., Sch\"{o}nfelder, V., Mandrou, P., et al.: 2003 
Atti\'e, D., Cordier B., Gros, M., Laurent, Ph., et al. 2003, Astronomy and
Astrophysics, 411, L63

\bibitem{verstrand}
Verstrand W.T., Share G.H., Murphy, R.J., Forrest D.J., Chupp, E.L., \&
Kanbach, G. 1999, Astrophysical Journal Suppl. Ser., 120, 409

\bibitem{weidens}
Weidenspointner, G., Harris,  M. J., Sturner, S., Teegarden, B.J.,  \& Ferguson
C. 2005, Astrophysical Journal Suppl. Ser. 156, 69

\bibitem{weiden2}
Weidenspointner, G., Kiener, J., Gros, M., et al. 2003, Astronomy and
Astrophysics, 411, L113 

\bibitem{werntz} 
Werntz, C., Lang, F.L., \& Kim, Y.E. 1990, Astrophysical Journal Suppl. Ser.
 73, 349

\end{thebibliography}
\end{document}